\documentclass[%
aps,
reprint,
superscriptaddress,
 amsmath,amssymb,
pra,
]{revtex4-1}

\usepackage{textcomp}
\usepackage[utf8]{inputenc}
\usepackage{gensymb}
\usepackage{graphicx}
\usepackage{dcolumn}
\usepackage{bm}
\usepackage{mathtools}
\usepackage{epstopdf} 
\usepackage{hyperref}
\usepackage{color}

\newcommand{\ket}[1]{| #1 \rangle}
\newcommand{\bra}[1]{\langle #1 |}

\begin{document}
\title{
Quantum Monte Carlo studies of a trimer scaling function with microscopic two- and three-body interactions 
}
\author{Lucas Madeira}
\email{madeira@ifsc.usp.br}
\affiliation{Instituto de F\'isica de S\~ao Carlos, Universidade de S\~ao Paulo, CP 369, 
13560-970 S\~ao Carlos, S\~ao Paulo, Brazil}

\author{Tobias Frederico}
\affiliation{Instituto Tecnol\'ogico de Aeron\'autica, DCTA, 12228-900 S\~ao Jos\'e dos Campos, Brazil}

\author{Stefano Gandolfi}
\affiliation{Theoretical Division, Los Alamos National Laboratory,
Los Alamos, New Mexico 87545, USA}

\author{Lauro Tomio}
\affiliation{Instituto de F\'isica Te\'orica, Universidade Estadual Paulista, 01405-900 
S\~ao Paulo, Brazil}

\author{Marcelo T. Yamashita}
\affiliation{Instituto de F\'isica Te\'orica, Universidade Estadual Paulista, 01405-900 
S\~ao Paulo, Brazil}

\date{\today}

\begin{abstract}
We present an energy scaling function to predict, in a specific range, the energy of bosonic trimers with large scattering lengths and finite range interactions, which is validated by quantum Monte Carlo calculations using microscopic Hamiltonians with two- and three-body potentials. The proposed scaling function depends on the scattering length, effective range, and a reference energy, which we chose as the trimer energy at unitarity. We obtained the scaling function as a limit cycle from the solution of the renormalized zero-range model with effective range corrections. We proposed a simple parametrization of the energy scaling function. Besides the intrinsic interest in theoretical and experimental investigations, this scaling function allows one to probe Efimov physics with only the trimer ground states, which may open opportunities to identify Efimov trimers whenever access to excited states is limited.
\end{abstract}

\maketitle

\section{Introduction}

Loosely bound few-body quantum systems close to unitarity, when the two-body scattering length diverges, have the corresponding densities distributed over a much larger region beyond the range of their mutual interactions. In this situation, the details of the inter-particle potential become almost irrelevant once the interaction can reproduce the ground-state spectrum. These basic quantum properties are fundamental in understanding few-body effects such as the Thomas collapse~\cite{Thomas1935} or the related well-known Efimov effect~\cite{Efimov1970,Efimov1981}. The former corresponds to the collapse of the three-body ground state when the range of the two-body interaction goes to zero. The latter tells us that an infinite number of three-body bound states can be found in the exact unitary limit. Later it was shown that both phenomena are closely related by a scale transformation~\cite{Adhikari1988}.

Bound and resonant states emerge as we approach the unitary limit. The independence on the details of the two-body potential was also observed by Phillips~\cite{Phillips1968} when studying the correlation between the triton binding energy and the doublet nucleon-deuteron scattering length by using several nucleon-nucleon potential models. Efimov and Tkachenko later explained this effect within the framework of the zero-range theory~\cite{EfimovTkachenko1985}.

The Efimov effect is also present in $N> 3$ systems, where $N$ is the number of particles. For example, Tjon studied the fixed-slope correlation between tetramer and trimer binding energies of $^4$He~\cite{Tjon1975}, the so-called Tjon line~\cite{Tomio2013}, which is closely related to the three-body Efimov physics~\cite{Platter2004,Platter2005,Yamashita2006,Ferlaino2009}. Coester \textit{et al.}~\cite{Coester1970} studied nuclear-matter binding energy variations with phase-shift-equivalent two-body potentials.

The study of few-nucleon correlations, which contributed to the characterization of the Thomas collapse and Efimov effect, followed pioneering mathematical approaches to three-body problems, done by  Skorniakov and Ter-Martirosian~\cite{Skorniakov1957}, Danilov~\cite{Danilov1961}, and Faddeev~\cite{Faddeev1965}. The search for three-body systems where Efimov states could be characterized began with few-nucleon and few-atom systems~\cite{Fonseca1979,Lim1980,Adhikari1982,Adhikari1982b}.

The experimental searches for Efimov states in nuclear physics encountered apparent limitations due to the properties of nucleon-nucleon interaction. The discovery of exotic nuclear systems, which could be described as a core with a two-neutron halo~\cite{Hansen1995b,Hansen1995} is the appropriate place in nuclear physics for investigations of possible bound or resonant states with Efimov character~\cite{Fedorov1994,Amorim1997}.

The first prediction of Efimov states in few-atomic systems was made by Lim \textit{et al.}~\cite{Lim1977}, pointing out its possible relevance in helium gases at low temperatures. This prediction was later made more convincingly when considering the possibility of one excited Efimov state in the three-helium atomic system~\cite{Cornelius1986}. For a review on the initial studies considering the Efimov nature of $^4$He trimer, the reader is referred to Ref.~\cite{Kolganova2011}. The excited Efimov state in $^4$He trimer was confirmed experimentally in 2015 by Kunitski \textit{et al.}~\cite{Kunitski2015}. The ultracold collision properties of $^4$He trimer have been discussed since 1997~\cite{Kolganova1997}, with access to new experimental data motivating investigations, in the context of Efimov physics, on the ultracold collision properties of a $^4$He dimer with a third atomic particle ($^4$He, $^{6,7}$Li, and $^{23}$Na)~\cite{Suno2017,Shalchi2020}.

The experimental realization of Bose-Einstein condensation with extremely diluted atom clouds in 1995~\cite{Anderson1995,Davis1995,Bradley1995} was a milestone for atomic physics. The possibility to control atom-atom interactions via Feshbach resonances~\cite{Feshbach1958,Feshbach1962,Timmermans1999,Chin2010} enabled the experimental search for Efimov states in atomic systems. Three-body Efimov states have been identified in different experimental setups, with the first evidence observed in an ultracold gas of cesium atoms~\cite{Kraemer2006}. The success in observing the existence of Efimov states in cold-atom systems is being explored by theoretical and experimental investigations~\cite{Braaten2006,Braaten2007}.

The spectrum of Efimov states in the limit of infinite scattering length is characterized by an asymptotic discrete scaling symmetry, which is the signature of renormalization group flow to a limit cycle. The connection between the Efimov effect and renormalization group limit cycles was noted in Ref.~\cite{Albeverio1981}. The Efimov effect was realized as a universal scaling limit of the three-body system~\cite{Amorim1997,Frederico1999} and within the framework of effective field theory (EFT)~\cite{Bedaque1999,Bedaque1999b}. A summary of the identified scales and universal aspects in few-body systems can be found in several reviews, considering nuclear and atomic systems~\cite{Frederico2011,Frederico2012a,Hammer2013,Zinner2013,Naidon2017,Greene2017,Kievsky2021}. The main relevant interest in the theoretical approaches is to establish possible extensions of the observed universality by considering range corrections, scaling limits, and extensions of such correlations to more than three particles, which can be observed in experimental investigations.

In the present work, our goal is to establish range corrections when considering a trimer, composed of identical bosons of mass $m$ close to unitarity, within the representation of a universal scaling function, which correlates the trimer energies for different scattering lengths with the effective range. This provides a practical framework yet untouched by previous related studies, such that extensions to larger systems can be followed systematically whenever possible. Linear range corrections to the zero-range approach were first considered to explain the Phillips line~\cite{EfimovTkachenko1985}. Since then, range corrections have been considered in several systematic studies~\cite{Efimov1991,
Hammer2001,Thogersen2008,Platter2009,Pricoupenko2010,Braaten2011,Castin2011,Werner2012,Wang2012,Tusnski2013,Gattobigio2014,Kievsky2014,Kievsky2015,Ji2015,Rodriguez2016,Souza2016}. In support to ongoing experimental investigations in cold-atom physics~\cite{Wild2012,Makotyn2014,Chapurin2019,Xie2020,Zou2021}, one can observe the interest in defining more precisely the expansion parameters near the unitary regime~\cite{Gattobigio2019,Mestrom2019}.

Recently, a quantum Monte Carlo (QMC) study up to $N=60$ bosons~\cite{Carlson2017} obtained the ground-state binding energies at unitarity for clusters with sizes much larger than the interaction range. Particularly, it is desirable to follow systematically, at least for the simplest non-trivial case of the three-boson system, the route of the energies when the unitarity limit is approached in terms of the scattering length, effective range, and a three-body scale. In this paper we propose an energy scaling function containing information about the two-body system, i.e., the scattering length and effective range, and the trimer energies at unitarity and finite scattering lengths. We investigated the scaling function, obtained as a limit cycle from the range-corrected Skorniakov and Ter-Martirosian equations, using QMC calculations with two- and three-body potentials for a specific range of the physical parameters mentioned above.

This work is structured as follows. In Sec.~\ref{sec:scaling} we introduce the basic concepts necessary to construct the energy scaling function and also the subtracted form of the Skorniakov and Ter-Martirosian equation with finite range corrections, which allows us to obtain numerically the limit cycle which characterizes the scaling function. The method employed in this work to study the trimer ground-state energy with finite range potentials, namely the quantum Monte Carlo approach, is presented in Sec.~\ref{sec:qmc}. We provide calculations with the Skorniakov and Ter-Martirosian zero-range model with finite range corrections to show the limit cycle of the proposed scaling function in Sec.~\ref{sec:energy}. In Sec.~\ref{sec:results}, we performed QMC simulations with microscopic two- and three-body potentials to compute the relevant quantities.  By comparing the results obtained with both methods, we select among the hundreds of systematical calculations the ones that are within the universal window, which allows us to characterize in detail the route toward unitary with finite effective ranges. Finally, our conclusions are presented in Sec.~\ref{sec:conclusions}. In the Appendix, we show that a zero-range parametrization of the Efimov states with an effective range function that runs with the energy scale motivates our ansatz for the scaling function.

\section{Universal scalings}
\label{sec:scaling}

\subsection{Effective range expansion}
\label{sec:dimer}

We start by reviewing some basic concepts applied to low-energy scattering and weakly bound states of two identical bosons. Let us consider the two-body scattering in the $s$-wave, described by the Schr\"odinger equation, with a finite range spherically symmetric potential $V(r)$, which depends only on the distance $r$ between the particles. The solutions are separated into radial $U(r)$ and angular $Y(\theta,\phi)$ parts, the latter being a constant for $s$-wave scattering. By defining $u(r)=r U(r)$, the two-body equation for particles with mass $m$ is given by
\begin{equation}
\label{eq:scattering}
\left[-\frac{\hbar^2}{2m_r}\frac{d^2}{dr^2}+V(r)\right]u(r)=\frac{\hbar^2 k^2}{2m_r}u(r),
\end{equation}
where $m_r=m/2$ is the reduced mass, and $\hbar^2 k^2/(2m_r)$ is the scattering energy. The scattering length $a$, which describes low-energy scattering, can be determined from the $k\to 0$ solution of Eq.~(\ref{eq:scattering}), $u_0(r)$,
\begin{equation}
\frac{1}{u_0(R)}\frac{d}{dr}u_0(r)\big|_R=\frac{1}{R-a},
\end{equation}
where $R$ is outside the potential range. Another relevant low-energy observable is the effective range $r_0$,
\begin{equation}
r_0 = 2 \int_0^\infty dr \left[\psi_0^2(r) -u_0^2(r) \right],
\end{equation}
where $\psi_0(r)$ is the asymptotic form of $u_0(r)$, that is,
the solution of Eq.~(\ref{eq:scattering}) with $k\to 0$ and $V(r)=0$.
The $s$-wave scattering length and corresponding effective range are related to the low-energy phase shift $\delta_0(k)$ through~\cite{Bethe1949}
\begin{equation}
\label{eq:delta}
k \cot \delta_0(k) = -\frac{1}{a}+\frac{r_0 k^2}{2} + \mathcal{O}(k^4).
\end{equation}
This equation is often a called shape-independent approximation because two different microscopic potentials 
that differ in shape produce the same low-energy phase shifts, as long as both have the same scattering 
length and effective range.
In systems where no three-body scale exists, such as two-component Fermi gases, 
Eq.~(\ref{eq:delta}) has allowed comparisons of 
results obtained with potentials of quite different shapes:
square-well \cite{Astrakharchik2004}, modified Poschl-Teller \cite{Carlson2003}, and even the $s$-wave
component of nuclear potentials \cite{Gezerlis2008,Gezerlis2010,Madeira2019}.

The two-body $s$-wave
scattering amplitude is given by
\begin{equation}\label{tau2}
    \tau(k)=\frac{1}{k\cot \delta_0-ik}\, .
\end{equation}
From the pole of the $s$-wave scattering amplitude in the first (second) complex energy sheet, one can
obtain the well-known scaling law for the bound (virtual) dimer energy $E_2=-\hbar^2/{(m a_B^2)}$, where
\begin{equation}\label{aB}
\frac{1}{a_B}=\frac{1}{a}-\frac 12 \frac{r_0}{a_B^2}\, .
\end{equation}
In this expression, we kept only the first two terms of the $k\cot\delta_0(k)$ effective range expansion.

\subsection{Trimer energy scaling}
\label{sec:trimer}
The three-boson system requires a three-body scale in the limit of a $s$-wave zero-range force to avoid the 
Thomas collapse, as the two-boson scattering length is not enough to determine the trimer low-energy 
properties. Corrections due to the effective range can be taken into account considering  the effective 
range expansion of Eq.~(\ref{eq:delta}) in the $s$-wave scattering amplitude.
If we choose a reference three-body energy at unitarity, $E_3({1}/{a}=0,r_0,\nu)$, where $\nu$ is a three-body scale, then it can be combined with the scattering length and effective range to produce 
two dimensionless quantities, which we denote by
\begin{eqnarray}
\label{eq:x}
x = \frac{\hbar}{a\sqrt{-m E_3(0,r_0,\nu)}},\\
\label{eq:y}
y = \frac{r_0\sqrt{-m E_3(0,r_0,\nu)}}{\hbar}\, ,
\end{eqnarray}
with these quantities defined such that $x=0$ corresponds to the unitary limit and $y=0$ to the zero-range limit.

Our goal is to establish an energy scaling function to obtain the trimer energy $E_3(1/a,r_0,\nu)$ as a function 
of the scattering length, effective range, and a three-body scale $\nu$.  As mentioned above,  
a reference energy is required, which we take to be the trimer energy at unitarity for a particular value 
of the effective range and three-body reference energy, $E_3(0,r_0,\nu)$. Given $x$ and $y$ by Eqs.~(\ref{eq:x}) and
(\ref{eq:y}), we define a scaling function $F(x,y)$ as
\begin{eqnarray}
\label{eq:universal}
F(x,y)\equiv\frac{E_3\left(1/a,r_0,\nu\right)}{E_3(0,r_0,\nu)}\, ,
\end{eqnarray}
which has to be determined, as well as the region where it displays a universal behavior, namely, where it does not depend explicitly on details of the microscopic interaction being considered for the examples we are going to explore.

The zero-range limit of Eq.~(\ref{eq:universal}) has been studied extensively in the literature, cast in a different form that contains the same information \cite{Braaten2006},
\begin{equation}
\label{eq:braaten0}
E_3(1/a,0,\nu) + \frac{\hbar^2}{2ma^2}= E_3(0,0,\nu)\exp{\left[\Delta(\xi)/s_0\right]},
\end{equation}
where $s_0$ is the Efimov parameter and
\begin{equation}
\tan \xi=-\left(\frac{m|E_3(1/a,0,\nu)|}{\hbar^2}\right)^{1/2} a.
\end{equation}
The $\Delta$ function, often called Efimov's universal function, can be determined by computing the binding energies in Eq.~(\ref{eq:braaten0}), which can be done with remarkable precision~\cite{Braaten2006,Naidon2017,Mohr2006,Gattobigio2019b}.
Our goal is to go beyond the zero-range limit and to compute trimer energies with finite effective ranges.

Although deriving an analytic expression for the scaling function of Eq.~(\ref{eq:universal}) is  
challenging, some of its features are known. Since the reference energies are calculated at unitarity,
$F(0,y)=1$ for all values of $y$. If we consider an expansion in powers of $x$ and $y$, a 
consequence is that every power of $y$ must be multiplied by a power of $x$. Also, the trimer 
energy for a finite scattering length and a zero-range interaction was computed up to first order in $1/a$ 
in the context of absorptive short-range potentials~\cite{Amorim1992}, and later in Refs.~\cite{Castin2011,Werner2012}, where it was shown that
\begin{equation}
\label{eq:Castin}
 \frac{\partial E_3(1/a,r_0,\nu)}{\partial(1/a)}=-\frac{\hbar^2 C_2}{8\pi m}.
\end{equation}
It is understood that the derivative must be taken at a fixed three-body scale,
and $C_2$ is the two-body contact.
Reference~\cite{Amorim1992} provides the value of the derivative 
in Eq.~\eqref{eq:Castin}, with the value $C_2=53.097\sqrt{-mE_3(0,r_0,\nu)}/\hbar$ being 
provided in Refs.~\cite{Castin2011,Werner2012}. By casting Eq.~(\ref{eq:Castin}) in the form of our 
scaling function, the zero-range behavior (for $|x|\ll 1$) emerges as 
\begin{equation}\label{eq:castinF}
F(x,0)=1+\left(\frac{53.097}{8\pi}\right)x\approx 1+2.113x.
\end{equation}
We should note that the zero-range assumption in deriving Eq.~(\ref{eq:Castin}) implies 
that we should only observe the linear behavior in $x$ with a slope of 2.113 if the reference 
energy $E_3(0,r_0,\nu)$ is computed at $r_0=0$. For finite values of the effective range, we expect the 
slope to be close to this value since both the numerator and denominator of Eq.~(\ref{eq:universal}) 
contain the finite $r_0$ dependence. To obtain the scaling function we will use the solutions of the 
Skorniakov and Ter-Martirosian (STM)~\cite{Skorniakov1957} equation with leading order effective 
range corrections.

\subsubsection{Skorniakov and Ter-Martirosian formalism}
\label{sec:stm}

The STM approach~\cite{Skorniakov1957} is the appropriate formalism for an analytical study of three-body systems close to the unitary limit, where universal aspects are dominant, not being affected by the details of two-body interactions. In Ref.~\cite{Amorim1992}, within a study considering a three-boson system with an absorptive short-range two-body potential, the STM formalism was applied in the limit of zero-range interaction, using a momentum cutoff renormalization.
From the linearity of an expansion of the STM equation, it was established in that work the universal constant 2.1 [which appears in Eq.~\eqref{eq:castinF} with two more digits]
for the first-order correction to the unitarity in the zero-range limit. 
By taking advantage of previous studies on the subtractive renormalization approach~\cite{Adhikari1995}, together with renormalization group invariance of quantum mechanics~\cite{Frederico2000}, used in the context of three-body scaling limit~\cite{Yamashita2002}, a subtracted form of the STM equation for bosonic trimer bound states with zero-range potential was found to be appropriate for a unified description of the Efimov and Thomas effects with cutoff regularization~\cite{Adhikari1988}.
With units such that $\hbar=m=1$, the $s$-wave subtracted STM equation for the trimer energy $E_3$  
can be written as~\cite{Frederico2012a}
\begin{multline}
f(q)=-\frac{2}{\pi} \tau\left(k\right)
\int_0^\infty dp \ p^2\, f(p)\int_{-1}^{+1}dz\, 
\\ \times
\left[ G_0(p,q,z;E_3)-G_0(p,q,z;-\nu)\right], \label{stm}
\end{multline}
where $k=i\sqrt{E_{3}-3 q^2/4}$ and the three-body Green's function is
$G_0(p,q,z;E)\equiv \left[E-p^2-q^2-p q z\right]^{-1}$,
with $\nu$ being a three-body short-range regularization parameter, presented as an
energy subtraction point in the formalism.
The two-body $s$-wave scattering amplitude is given by Eq.~\eqref{tau2}, which 
in the lowest order of the effective range $r_0$ leads to \cite{Hadizadeh2013}
\begin{equation}\label{eq:tau}
 \tau(k)=\left({-\frac{1}{a_B} -ik}\right)^{-1}
 \left[1+ \frac{r_0}{2}\left(\frac{1}{a_B}-ik\right)\right],
\end{equation}
with $a_B$ related to the 
scattering length 
by Eq.~\eqref{aB}.

\section{Quantum Monte Carlo methods}
\label{sec:qmc}

\subsection{The microscopic Hamiltonian}

The Hamiltonian we considered for the three identical bosons is given by
\begin{equation}
\label{eq:H}
H = -\frac{\hbar^2}{2m}\sum_{i=1}^3 \nabla^2_i+\sum_{i<j} V_2(r_{ij})+V_3(R_{123}),
\end{equation}
where $r_{ij}=|\bm{r}_i - \bm{r}_j|$ are the pair distances, with $V_2(r_{ij})$ the 
corresponding pairwise attractive interactions.
The term $V_3(R_{123})$ is a repulsive three-body 
force, where $R_{123}\equiv(r_{12}^2+r_{13}^2+r_{23}^2)^{1/2}$.

Once we consider that the function $F(x,y)$ defined in Eq.~(\ref{eq:universal}) is universal, then it must be independent of the microscopic details of the interactions. To verify such universality of $F(x,y)$ first we compare our results with the ones obtained using the subtracted STM equation. Then, for some selected values of $x$ and $y$, we employed two different two-body interactions tuned to reproduce the desired values of the scattering length and effective range.

The majority of our results is obtained with a Gaussian potential, defined by
\begin{equation}
\label{eq:gauss}
V_{\rm G}(r)=-\lambda_{\rm G} \frac{\hbar^2 \mu_{\rm G}^2}{m_r} 
\exp\left[-\frac{\mu_{\rm G}^2r^2}{2}\right],
\end{equation}
with the tunable parameters $\lambda_{\rm G}$ and $\mu_{\rm G}$.
This potential is frequently employed in the cold atom gases literature~\cite{Carlson2017}.
Another choice was the modified Poschl-Teller potential, which has been successfully used to 
describe interactions in cold atom systems \cite{Carlson2003,Madeira2016,Madeira2017,Madeira2019}. It is 
given by
\begin{equation}
\label{eq:mPT}
V_{\rm mPT}(r)=-\lambda_{\rm PT} \frac{\hbar^2 \mu_{\rm PT}^2}{m_r} \frac{1}{\cosh^2(\mu_{\rm PT} r)}.
\end{equation}
This potential is commonly employed for convenience since there is
an analytical expression that relates the parameters $\lambda_{\rm PT}$,  
$\mu_{\rm PT}$, and $a$~\cite{Madeira2019}.

Although both potentials share some similarities, such as having two tunable parameters and representing smeared out delta functions, they produce different results for the same values of $a$ and $r_0$. 
For example,  in the absence of a three-body force, the trimer energy at unitarity is
$E_3 m r_0^2/\hbar^2= -0.54(1)$ if computed with the modified Poschl-Teller potential and 
$-0.49(1)$ by using the Gaussian potential.
Hereafter, we report our results with the uncertainty of one standard deviation.
These differences persist even for other 
different scattering lengths, effective ranges, and non-zero three-body forces. 
Hence, they represent sensible choices in our aim for testing the universal scaling of 
Eq.~(\ref{eq:universal}).

For the three-body force in Eq.~(\ref{eq:H}), the following Gaussian form~\cite{Carlson2017} was chosen: 
\begin{equation}
\label{eq:V3}
V_3(R_{123})=\lambda_3 \frac{\hbar^2 \mu_3^2}{m} 
\exp\left[-\frac{\mu_3^2 R_{123}^2}{2}\right].
\end{equation}
The dimensionless strength parameter $\lambda_3$ is increased to produce loosely bound trimers, while the parameter 
$\mu_3$ controls the range of the interaction.

In physical systems, the expected geometric tower of Efimov states at unitarity is truncated 
from below due to the finite interaction range. The binding energy of the next deeper trimer would be
$\approx 22.7^2$ larger than the ground-state one.
Hence, it has been argued that if the potential parameters are such that $\hbar\mu_{2,3}/\sqrt{-m E_3} \gg 22.7$,
then the shape of the potential should produce small effects~\cite{Bedaque1999b,Bedaque1999,Carlson2017}.
However, in our investigation we also focus on cases where $\hbar\mu_{2,3}/\sqrt{-m E_3} < 22.7$ and finite range corrections are appreciable.

\subsection{Diffusion Monte Carlo method}

The solution of the Schr\"odinger equation using
two- and three-body finite-range potentials is obtained with variational and diffusion 
Monte Carlo methods, called VMC and DMC, respectively.
The VMC method relies on a trial wave function which should capture the properties of 
the ground state. The trial wave function contains variational parameters, which are optimized to 
minimize the energy.

Universality tests of bosonic trimers using pairwise interactions often employ trial wave functions consisting of a zero-range part multiplied by two-body correlations. In Ref.~\cite{Naidon2014}, for example, the authors constructed a trial wave function for the van der Waals potential by multiplying the uncorrelated hyperangular expression of the Efimov trimer in the zero-range limit~\cite{Efimov1971} by the zero-energy two-body wave function~\cite{Gao1998,Flambaum1999}. This variational approach was successful in verifying their interpretation of the microscopic origin of the universal three-body parameter.

Since the Hamiltonian, Eq.~(\ref{eq:H}), contains a three-body force, the inclusion of three-body correlations in our trial wave functions greatly reduces both the variational energy and the variance of the results.
The considered trial wave functions are of the form \cite{Carlson2017}
\begin{equation}
\label{eq:psit}
\psi_T(\mathcal{R})= \left( \prod_{i=1}^3 f_1(r_i)\right) \left(\prod_{i<j} f_2(r_{ij})\right) f_3(R_{123}),
\end{equation}
where $\mathcal{R}$ is a shorthand notation for all the coordinates.
The one-body factor $f_1(r)=\exp(-\alpha r^2)$ describes the cluster formation, where we 
introduced the variational parameter $\alpha$, and the distance $r_i$ is taken from the center of 
mass of the cluster.
The short-range correlations are included via a two-body Jastrow function $f_2(r)$.
Its boundary conditions are $f_2(r>d)=1$ and $f_2'(d)=0$ (where $d$ is a variational parameter often called 
healing length).
The zero-energy solution of the Schr\"odinger equation with the modified Poschl-Teller potential, 
Eq.~(\ref{eq:mPT}), is proportional to $\tanh(\mu_{\rm PT} r)/r$,
thus we adopted the form $f_2(r)=K\tanh(\mu_J r)\cosh(\gamma r)/r$ (for $r\leqslant d$),
where $K$ and $\gamma$ are adjusted to match the boundary conditions, with $\mu_J$ being 
a variational parameter.

The three-body factor describes the effect of the three-body repulsive force,
$f_3(R)=\exp\left\{u_0\exp[-R^2/(2R_0^2)]\right\}$, where $u_0$ and $R_0$ are determined at the VMC level. 
For a given two-body potential (either the modified Poschl-Teller or Gaussian) and a three-body force, all the variational parameters are optimized 
by using VMC simulations. Then, the optimized trial wave function is used as input in the DMC calculations.
Since we are dealing with a bosonic system, there is no sign problem, and thus the energies obtained with the DMC method are exact within statistical uncertainties. The quality of the variational wave function only reduces the variance and improves the convergence of the results.

The diffusion Monte Carlo method projects out the lowest-energy state of $H$, which is described by an initial state $\psi_T$, Eq.~(\ref{eq:psit}), by propagation in the imaginary time $\tau$,
\begin{equation}
\psi(\tau)=\exp\left[-(H-E_T)\tau\right]\psi_T,
\end{equation}
where $E_T$ is an energy offset. In the limit $\tau\to\infty$ only the lowest-energy state $\Phi_0$ 
survives, since higher-energy components will be exponentially damped.
The imaginary time evolution is governed by
\begin{equation}
\label{eq:green}
\psi(\mathcal{R},\tau)=\int d\mathcal{R}' \ G(\mathcal{R},\mathcal{R}',
\tau)\psi_T(\mathcal{R}'),
\end{equation}
where $G(\mathcal{R},\mathcal{R}',\tau)$ is the Green's function associated with the Hamiltonian.
The Green's function contains two terms: a diffusion term related to the kinetic energy operator, 
and a branching term related to the potential. We solve an importance sampled version \cite{Foulkes2001} of 
Eq.~(\ref{eq:green}) iteratively by using the Trotter-Suzuki approximation ~\cite{Trotter1959,Suzuki1976,Suzuki1985},
which requires the time steps $\Delta \tau$ to be small. For a detailed description of the DMC algorithm, 
the reader is directed to Ref.~\cite{Foulkes2001} and references therein.

Expectation values of operators that do not commute with the Hamiltonian, in our case the mean 
square radius, can be computed by using extrapolated estimators
\begin{equation}
\bra{\Phi_0} O \ket{\Phi_0} \approx \frac{\bra{\Phi_0} O \ket{\psi_T}^2}{\bra{\psi_T} O 
\ket{\psi_T}}+\mathcal{O}[(\Phi_0-\psi_T)^2],
\end{equation}
where the results of DMC and VMC runs are combined.

\section{Energy scaling function}
\label{sec:energy}

\subsection{Limit cycle and effective range}

We started our calculations by using the subtracted form of the STM equation, as given in Sec.~\ref{sec:stm}, to compute the trimer energies at unitarity for values of $y$ in the range $0 \leqslant y \leqslant 0.30$.
From Eq.~(\ref{eq:x}), we have that these energies correspond to $x=0$ in the scaling function. Then, we computed the trimer energies at other scattering lengths departing from unitarity. To make comparisons easier, we chose to compute all trimer energies for a fixed set of $x$ values, 
$-0.15\leqslant x \leqslant 0.15$ equally spaced by 0.05 increments.
Since the value of the reference energy varies, a given value of $x$ corresponds to different values of $a$ depending on $E_3(0,r_0,\nu)$, as shown by Eq.~(\ref{eq:x}).

\begin{figure}[htb]
\centering
\includegraphics[angle=-90,width=0.95\linewidth]{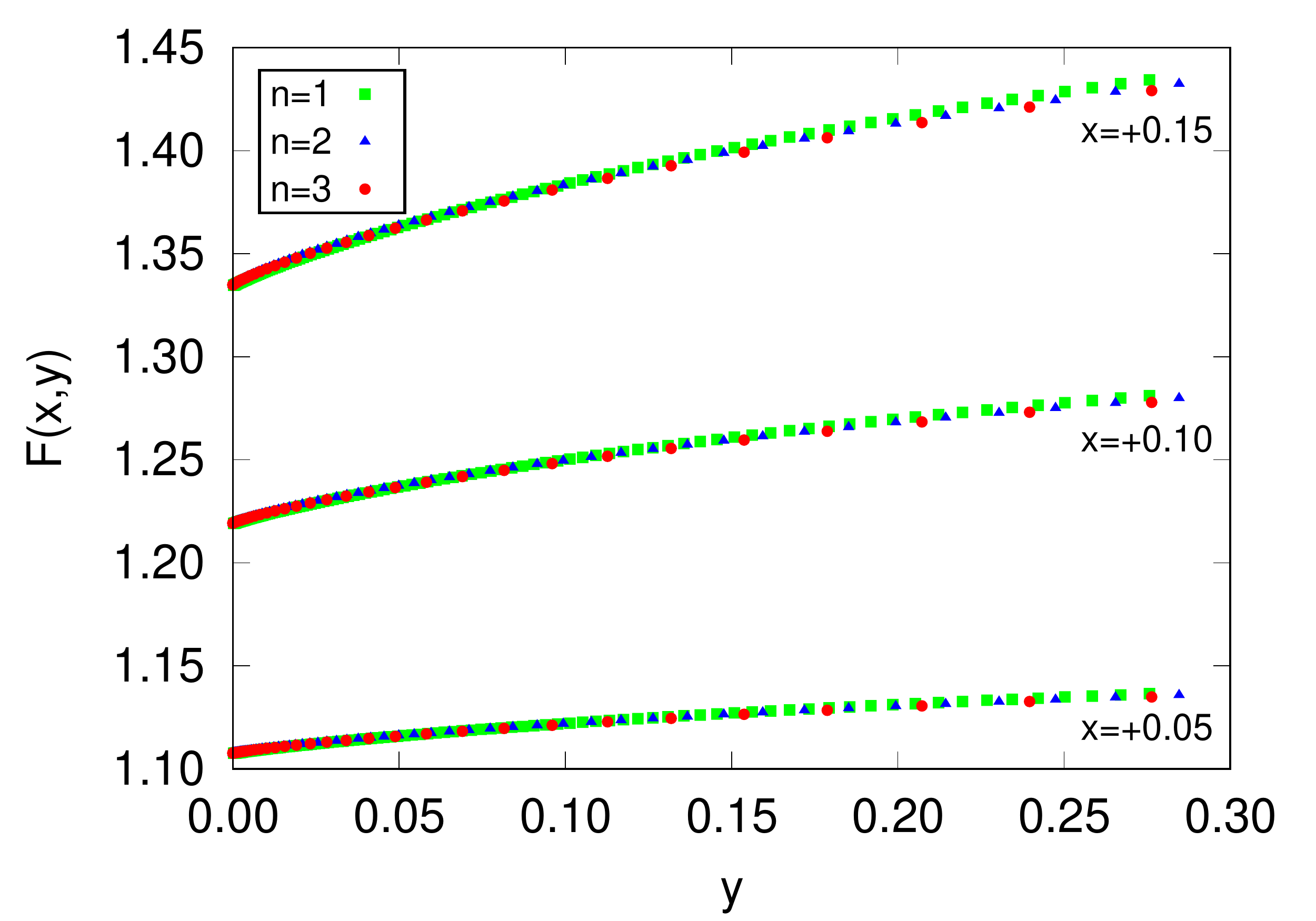}
\includegraphics[angle=-90,width=0.95\linewidth]{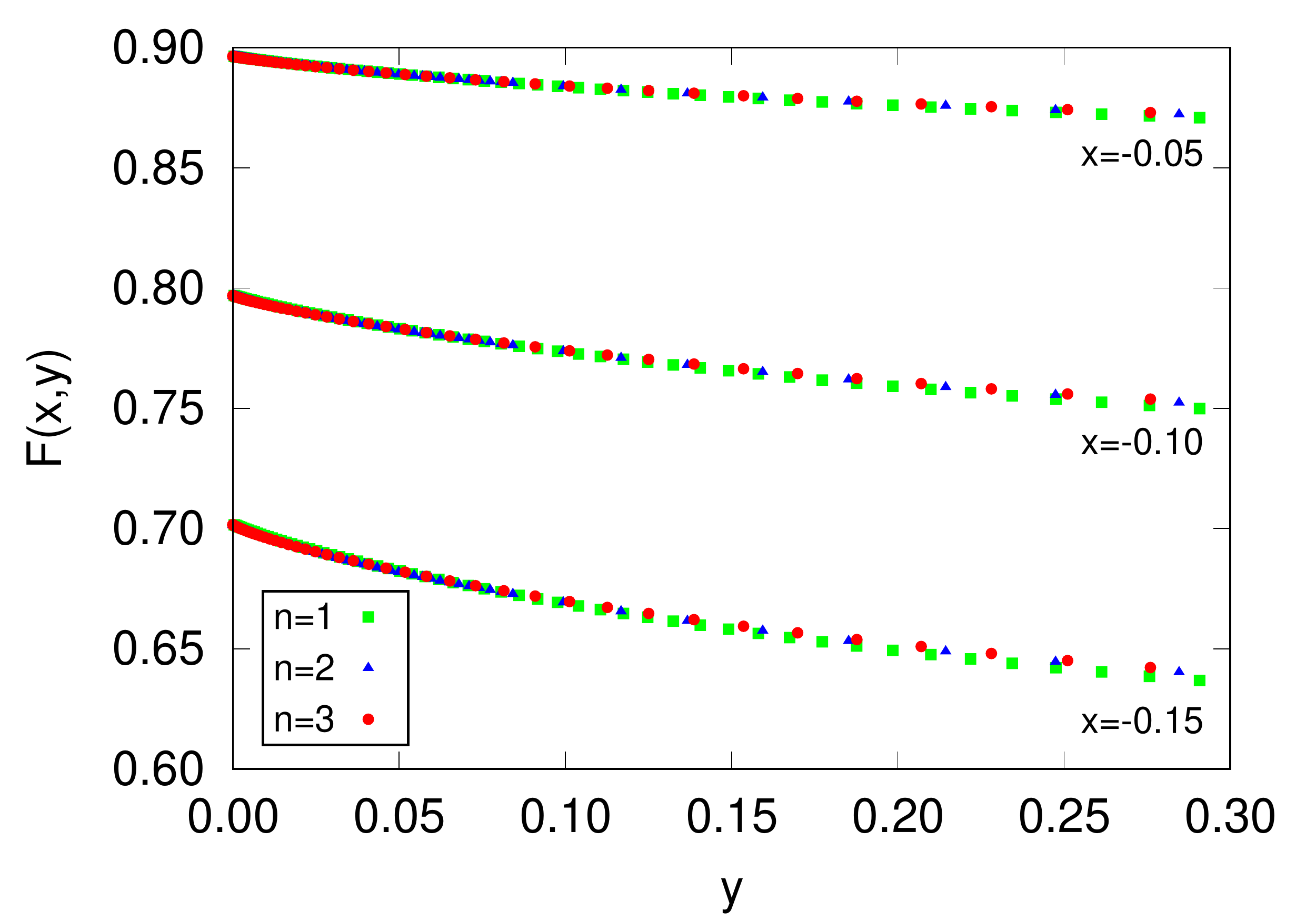}
\caption{(Color online) Limit cycle of the energy scaling function 
$F(x,y)=E_3(1/a,r_0,\nu)/E_3(0,r_0,\nu)$ as a function of
$y = r_0\sqrt{-m\, E_3(0,r_0,\nu)}/\hbar$. In the upper panel we present the results for 
$x = \hbar/(a\sqrt{-m \,E_3(0,r_0,\nu)}>0$, and the lower panel corresponds to $x<0$. 
The (green) squares, (blue) triangles, and (red) circles stand for the first, second, and third 
excited states, respectively.
}
\label{fig:limitcycle}
\end{figure}

We repeated the procedure described above for the first three excited states ($n=1$, 2, and 3) of the trimer. In Fig.~\ref{fig:limitcycle}, we show our results for the scaling function $F(x,y)$
given by Eq.~(\ref{eq:universal}), in terms of the variable $y$, which is related to the effective range $r_0$ by Eq.~(\ref{eq:y}).
We present the results for negative and positive scattering lengths ($x<0$ and $x>0$, respectively).
Although the results are essentially the same for small values of $y$, for larger values of this parameter, there is a slight difference in the slope of the curves between $n=1$ and the two following excited states. The size of the trimers increases by a factor of 22.7 if we compare the ground state to the first excited state or between two consecutive excited states. Hence, as we move toward higher values of $n$, the results are less sensitive to the regularization parameter. Since the function should have already reached a converged limit cycle for $n=3$, we employed the results obtained with the third excited state for the remainder of this work.

\subsection{Scaling function parameterization}

\begin{figure}[!htb]
\centering
\includegraphics[angle=-90,width=8.5cm]{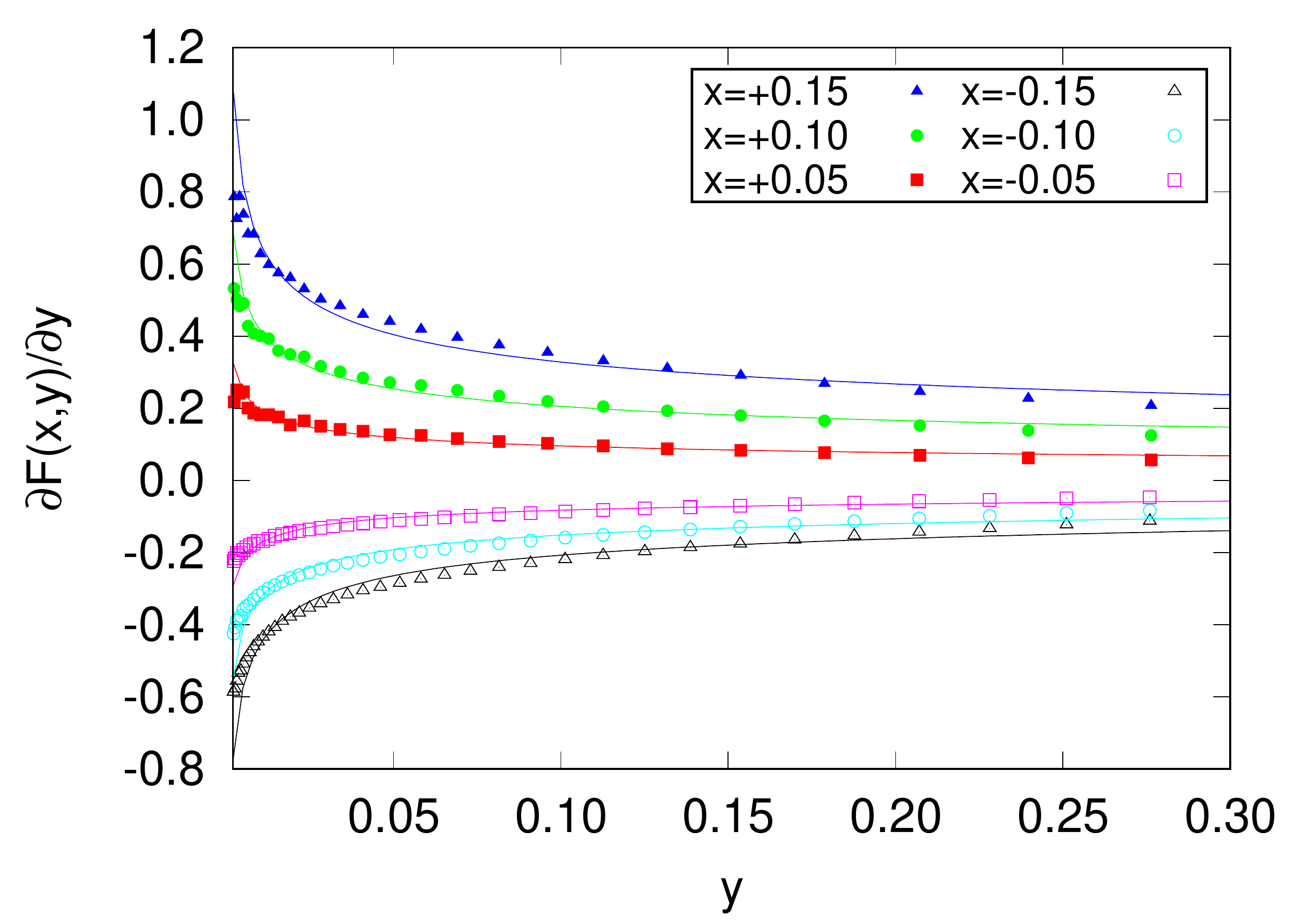}
\includegraphics[angle=-90,width=8.5cm]{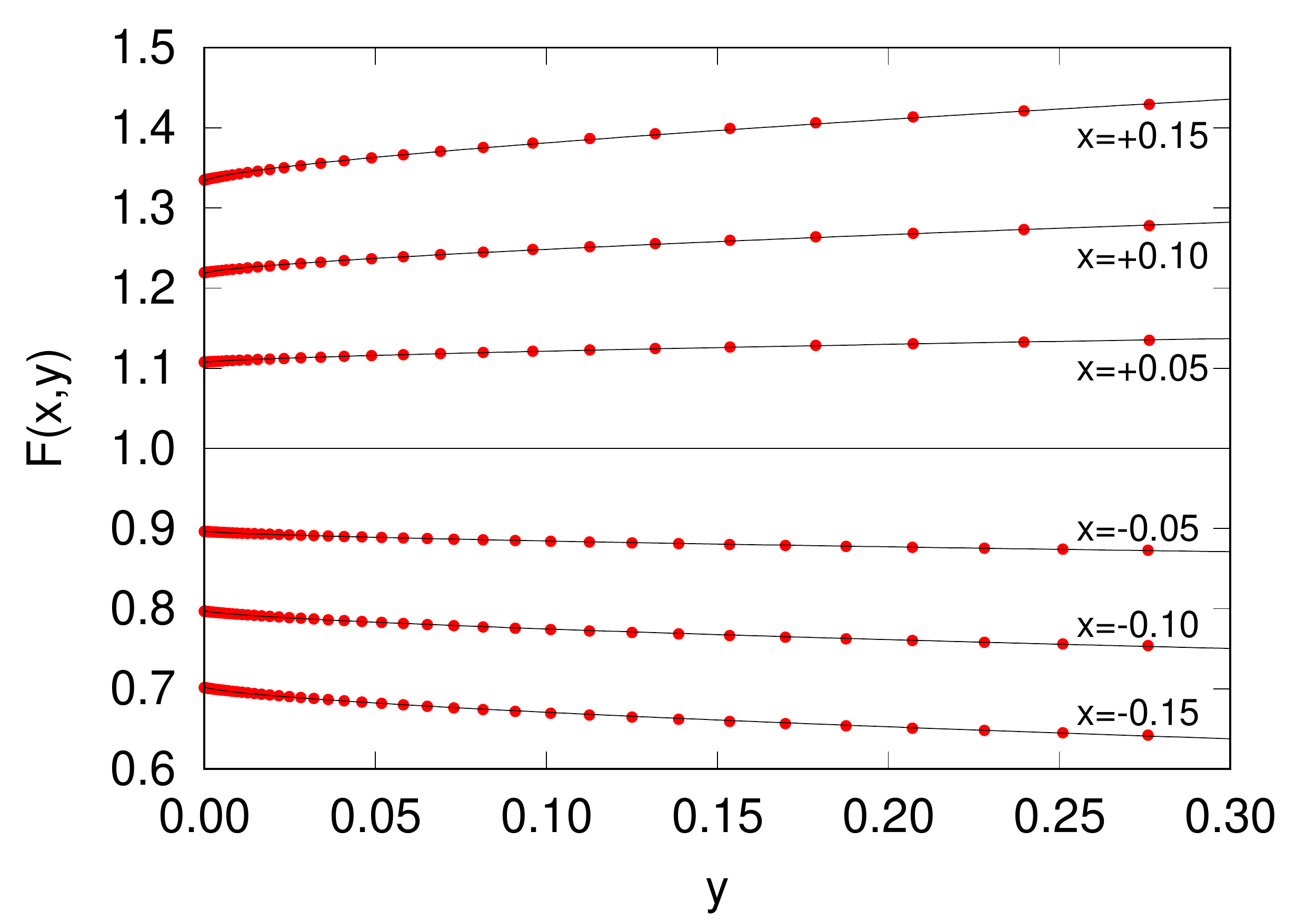}
\caption{(Color online) 
Upper panel: numerical derivative of $F(x,y)$ with respect to $y$, for the $n=3$ state shown in 
Fig.~\ref{fig:limitcycle}. Open (filled) symbols are for $x<0$ $(x>0)$. The solid curves correspond 
to the derivative of Eq.~(\ref{eq:F}). Lower panel: energy scaling function $F(x,y)$ fitted to the 
functional form of Eq.~(\ref{eq:F}). The (red) circles refer to results with the subtracted form of 
the STM equation with  finite-range corrections, while the curves correspond to the fit.
}
\label{fig:fit}
\end{figure}

The energy scaling function shows a curvature when approaching the zero-range limit, a behavior 
that one can verify by looking closely at the $y\to 0$ results plotted in Fig.~\ref{fig:limitcycle}. 
This effect can be made more visible by performing the numerical derivative of the scaling function 
with respect to $y$; see the upper panel of Fig.~\ref{fig:fit}.
To capture this feature of $\partial F(x,y)/\partial y$ close to the origin, we included terms 
proportional to $y^\sigma$, with $0<\sigma<1$, in the expansion of the scaling function.

Considering all the desirable features of $F(x,y)$ we presented so far, we proposed the 
following functional form for the scaling function: 
\begin{equation}
\label{eq:F}
F(x,y)=1+c_1 x+c_2 xy^\sigma+c_3 x^2+c_4 x^2y+c_5 x^2y^\sigma,
\end{equation}
which corresponds to an expansion in powers of $x$ and $y$. The parameters $c_i$ and $\sigma$ were 
determined by fitting the STM results to Eq.~(\ref{eq:F}), and we provide their values in 
Table~\ref{tab:1}. For reference, we show the function and its numerical derivative with respect 
to $y$ alongside the STM results in Fig.~\ref{fig:fit}. Besides having the desired features 
to describe both the energy function and its derivative, this functional form is further motivated 
in the Appendix.

\begin{table}[htb]
\caption{
Coefficients of Eq.~\eqref{eq:F} which corresponds to the functional form of $F(x,y)$. The first 
line represents the values obtained by fitting the equation with the STM results. The second line 
corresponds to Eq.~\eqref{fxy-1}, in which $\sigma=2s_0/\pi$, obtained with the ansatz described 
in Appendix~\ref{sec:appendix}. The uncertainties correspond to one standard deviation.}
    \label{tab:1}
    \centering
    \begin{tabular}{c c c c c c c }
    \hline\hline
  &  $c_1$ & $c_2$ & $c_3$ & $ c_4$ & $c_5$ & $\sigma$\\ \hline
Fit  &  2.106(1) & 1.26(1)&0.804(4)& 1.0(2)& 1.2(1)& 0.680(3) \\
Ansatz & 2.107    & 1.35 & 0.804 & 1.03 & 1.05 & 0.641 \\
 \hline\hline 
    \end{tabular}
\end{table}

To fit the limit cycle results for the energy scaling function, we rely on an ansatz, detailed in the Appendix, which gives
\begin{multline}\label{fxy-1}
F(x,y)=1+2.107\, x+1.35\,y^{\frac2\pi s_0}\,x\\   +\left(0.804+1.03\,y+1.05 \, s_0y^{\frac2\pi s_0}\right)x^2+\cdots .
\end{multline}
With this ansatz, the $\sigma$ exponent of $y$ suggested in Eq.~(\ref{eq:F}), $0<\sigma<1$, is related to the Efimov parameter $s_0$ by $\sigma=2s_0/\pi$. All the coefficients of this expression are defined in correspondence with the ones given in Eq.~(\ref{eq:F}). By comparing the coefficients obtained with the fitting procedure and the ones given by Eq.~(\ref{fxy-1}), we observe a remarkably close agreement between the expressions, as evidenced by the coefficients for both expansions shown in Table~\ref{tab:1}.

\section{Quantum Monte Carlo results}
\label{sec:results}

Once we have determined the energy scaling function from the limit cycle, obtained by solving the subtracted form of the STM equation with effective range corrections, our next task is to compute it with microscopic two- and three-body interactions. By comparing results obtained with both methods, we will consolidate the validity of the universal scaling approach to potential models.

To solve the Schr\"odinger equation with microscopic interactions, we chose to apply quantum Monte Carlo methods, as described in Sec.~\ref{sec:qmc}. Using these computational methods to obtain the trimer ground-state energies, we hope to provide a reliable view of the applicability of the universal scaling analysis emerging in three-body systems. Within this framework, one may extend the approach to realistic weakly bound few-body systems, such as those studied in cold atom laboratories.

We performed a large number of calculations with a fixed effective range ($\mu_2 r_0$=2) and varying the scattering lengths for Gaussian two- [Eq.~(\ref{eq:gauss})] and three-body [Eq.~\eqref{eq:V3}] interactions. For some selected values of $x$ and $y$ we employed the modified Poschl-Teller [Eq.~\eqref{eq:mPT}] two-body potential instead. Among such several QMC calculations, we will show that within a certain potential parameters window, it is possible to select results supporting the applicability of the universal energy scaling function. The analysis of this parameter window is also complemented by the calculation of the trimer mean square radius. 

\subsection{Systematic calculations}

First, we computed the trimer energies at unitarity, the denominator of Eq.~(\ref{eq:universal}). Varying the strength ($\lambda_3$) of the repulsive three-body force changes the trimer energy and, consequently, the value of $y$, Eq.~(\ref{eq:y}). From Eq.~(\ref{eq:x}), we have that all energies computed at unitarity correspond to $x=0$ in the scaling function. The ratio of the trimer energy at unitarity without a three-body force ($\lambda_3=0$) to the one with the most repulsive three-body force considered in this work is of the order of 1000.
We kept the value of the effective range fixed throughout all simulations, so for each three-body force at unitarity we have a different value of $y$, as seen in Eq.~(\ref{eq:y}). So far, the described procedure yields all the reference energies that we need to compute the denominator of the energy scaling function. The numerator comes from the trimer energies calculated out of the unitary limit for finite values of the two-body scattering length.

\begin{figure}[!htb]
\centering
\includegraphics[angle=-90,width=0.95\linewidth]{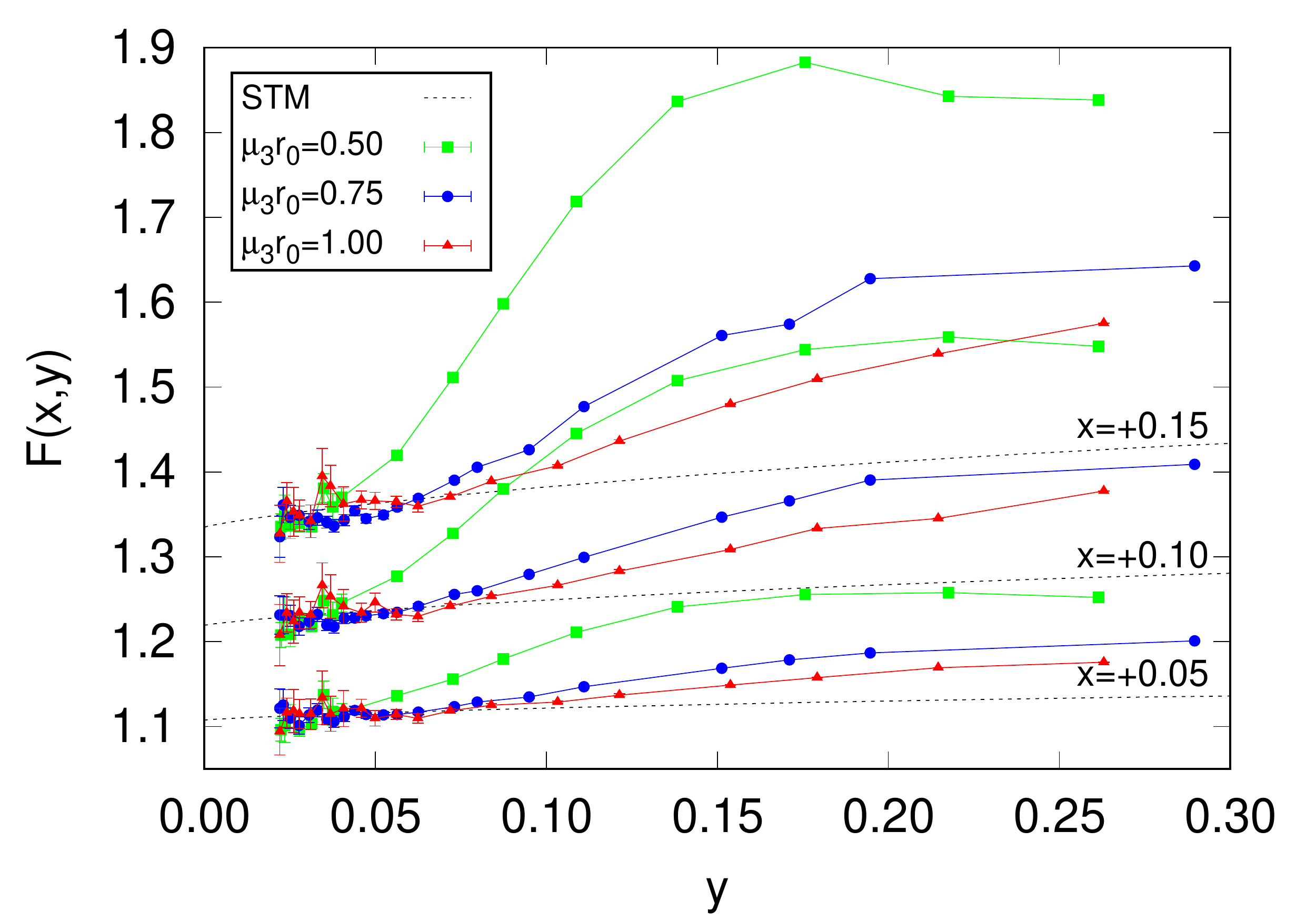}
\includegraphics[angle=-90,width=0.95\linewidth]{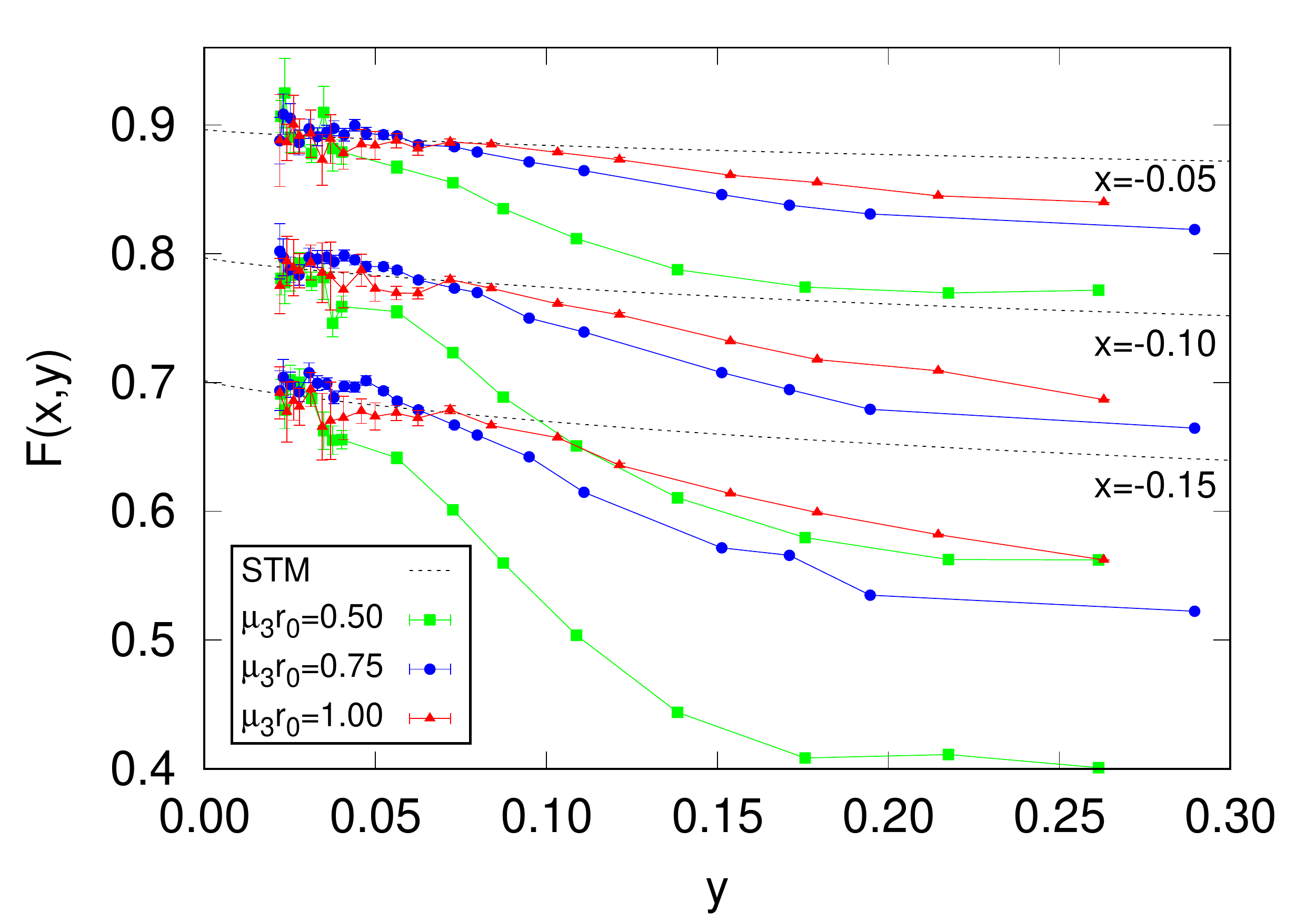}
\caption{(Color online) 
Energy scaling function calculated with QMC results (data points) compared with the one obtained from the solution of the STM subtracted equation with effective range corrections (dashed curves). The QMC results concern the trimer ground-state energies computed with microscopic Hamiltonian containing Gaussian two- and three-body potentials. The simulations were performed for three different ranges of the three-body force, $\mu_3r_0=$0.50, 0.75, and 1.00, (green) squares, (blue) circles, and (red) triangles, respectively. The range of the repulsive three-body potential is proportional to $1/\mu_3$,Eq.~(\ref{eq:V3}). Results for the same value of $x$ are connected by lines to guide the eye.
}
\label{fig:all}
\end{figure}

In Fig.~\ref{fig:all} we present our QMC results using Gaussian two- and three-body potentials alongside the scaling function obtained with the subtracted form of the STM equation. We employed three different values of $\mu_3$ in Eq.~(\ref{eq:V3}), $\mu_3r_0=$0.50, 0.75, 1.00, which is equivalent to varying the range of the three-body force (proportional to $1/\mu_3$).

For small values of $y$, the results using microscopic Hamiltonians coincide with the scaling function within statistical errors. As we move toward larger values of $y$, the results deviate from the scaling function. As expected, smaller values of $\mu_3$ correspond to larger ranges of the three-body force, which produce greater differences with the scaling function.

Our calculations are performed with $y\geqslant 0.02$ because of the statistical errors for smaller values of $y$, as explained in the following. The scaling function depends on the ratio of two energies, the trimer energy at some finite scattering length and at unitarity. As we move toward small values of $y$, with a fixed effective range, the trimers become very loosely bound and quite large. The statistical errors of our simulations also decrease when going to this limit, but not as fast as the binding energies. Hence, although simulations with $y<0.02$ are possible, they would produce substantial errors for the computation of the scaling function.

The behavior of the scaling function with $y$ is the same as seen in Fig.~\ref{fig:all}, namely, it increases with $y$ for $x>0$, and it decreases for $x<0$. The energy ratio for larger values of $y$ is obtained by decreasing the strength of the three-body force. In the absence of a three-body potential ($\lambda_3=0$) the trimer energy at unitarity depends only on the choice of the two-body potential. For the Gaussian potential, $E_3 m r_0^2/\hbar^2=-0.49(1)$, which corresponds to a value of $y=0.7$.

The significant variations of the scaling function $F(x,y)$ from the Gaussian two and three-body potentials with respect to the limit cycle seen in Fig.~\ref{fig:all}, can be understood as follows.
For $x>0$, the attractive two-body force is stronger than the one at the unitary limit, as reflected by the finite value of the two-body binding. This extra attraction compensates the repulsion of the three-body potential. By decreasing the magnitude of the repulsive three-body force, corresponding to increasing $y$, such an effect is stronger than the change of the on-shell scattering amplitude given by Eq.~\eqref{eq:tau}, when the effective range increases and leads to the observed limit cycle. 
For $x<0$, the attractive two-body force is weaker than the one in the unitary limit, which  weakens the trimer binding with respect to its value at unitarity. In the presence of the repulsive three-body force such effect is mitigated, and we observe the sharp increase of $F(x,y)$ as $y\to 0$ for the potential model results in comparison to the limit cycle ones.

\subsection{Universal window}
\label{sec:universal}

The results shown in Fig.~\ref{fig:all} indicate model dependence for large values of $y$, but they agree with the scaling function for small values of $y$. Hence, we would like to determine the conditions that yield the universal behavior described by the scaling function.

Trimer properties are expected to be universal if the ground-state energy fulfills two independent conditions: (i) $x=\hbar/(a\sqrt{-m E})\to 0$ and (ii) $y=r_0\sqrt{-m E}/\hbar\to 0$, characterizing a weakly bound trimer 
when the ratio $r_0/a \to 0$. We considered finite range interaction models where the effective
range is such that $r_0/|a|\lesssim 0.05$.  Condition (i)  can always be achieved going to the unitary limit, even for strongly bound trimers well within the potential range. These trimers do not bear universal properties. The associated probability of finding the bosons outside the potential range is small, contrary to what is expected from a universal Efimov trimer.
Condition (ii) is achieved  by tuning the repulsive short-range three-body potential, which makes the trimer ground-state  weakly bound. Universality is expected if the trimer is large with respect to the  ranges of the two- and three-body interactions.

\begin{figure}[!htb]
\centering
\includegraphics[angle=-90,width=\linewidth]{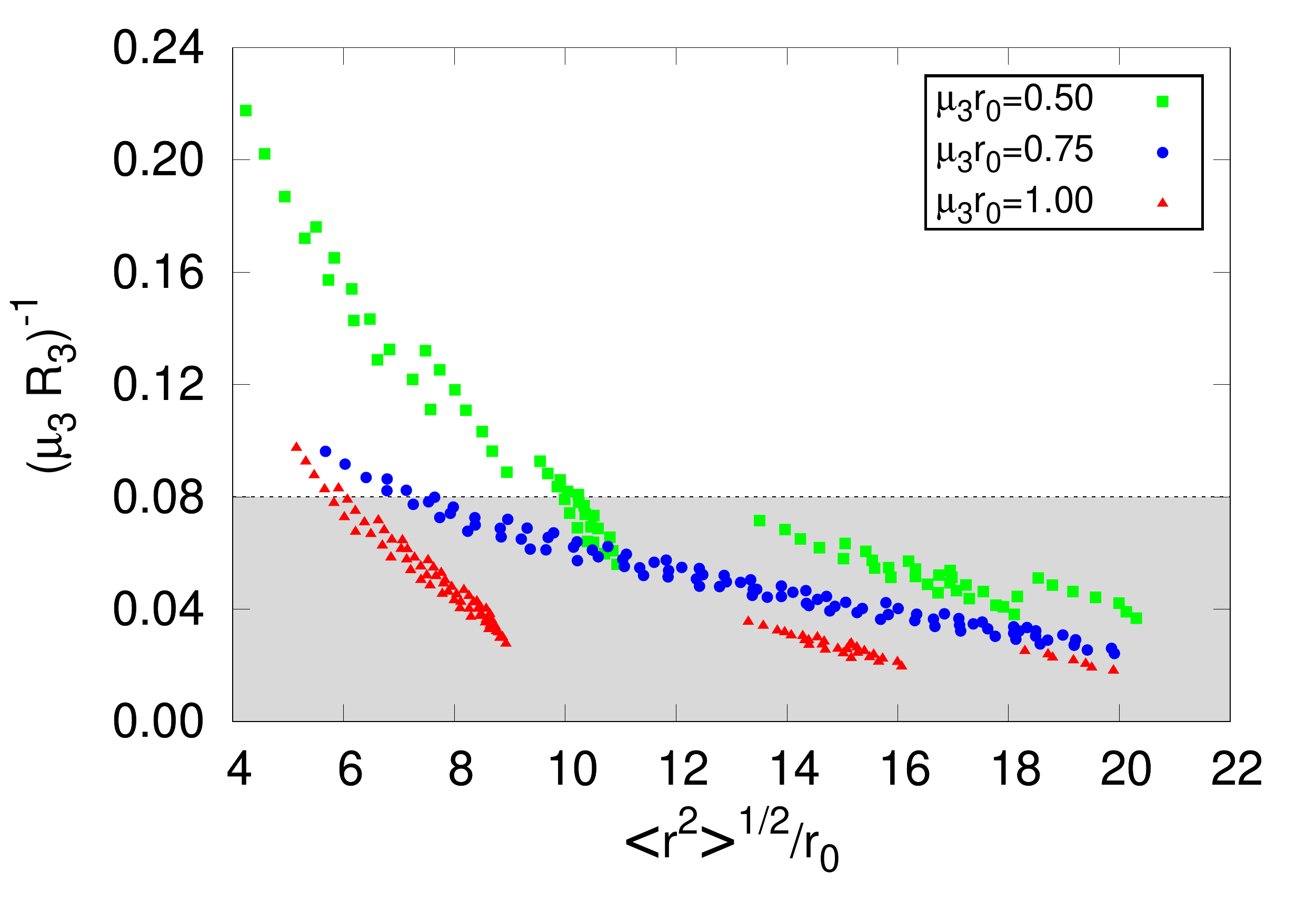}
\caption{(Color online) 
Plot of $1/(\mu_3 R_3)$ as a function of the square root of the mean square radius divided by the effective range. We only included points of Fig.~\ref{fig:all} with $y\leqslant 0.10$. The results shown are obtained with the Gaussian two-body potential and the repulsive Gaussian three-body force with different ranges, namely $\mu_3 r_0=$ 0.50, 0.75, and 1.00, (green) squares, (blue) circles, and (red) triangles, respectively.
}
\label{fig:mu3_r2}
\end{figure}

The quantity $R_3=\hbar/\sqrt{-m E}$ defines a characteristic length scale of the trimer. In Fig.~\ref{fig:mu3_r2} we plot $1/(\mu_3 R_3)$ as a function of the square root of the mean square radius over the effective range, $\sqrt{\langle r^2\rangle}/r_0$. Since $1/\mu_3$ is proportional to the range of the three-body force, the quantity in the ordinate axis measures the range of the three-body force compared to the size of the trimer. Universality is expected for $1/(\mu_3 R_3)\ll 1$ and $\sqrt{\langle r^2\rangle}/r_0\gg 1$, where the last condition also leads to $r_0\sqrt{-m E}/\hbar\ll 1$.

We found consistently that trimers with $1/(\mu_3 R_3)\leqslant 0.08$, the shaded region of the plot, agree with the scaling function, as we will show in the following subsection. Notice that a matching criterion considering the ratio of $\sqrt{\langle r^2\rangle}$, which also defines a typical length for the trimer, to the effective range would exclude points that are in agreement with the scaling function. For the universal trimers, the values of $\sqrt{\langle r^2\rangle}/r_0$ range from approximately 6 to 20. Again, this shows that universality for these potential models is much more sensitive to the range of the three-body force than to the trimer size.

A prediction of the Efimov nature of the $^4$He trimer was made in 1977~\cite{Lim1977}, but it was not experimentally observed until 2015~\cite{Kunitski2015}. This long period allowed for extensive investigations concerning the Efimov universality in helium systems~\cite{Kolganova2011}. \textit{Ab initio} variational calculations concerning few-body $^4$He systems have been performed with a plethora of realistic pairwise potentials: LM2M2 \cite{Aziz1991,Hiyama2012a}, a potential that takes into account relativistic and quantum electrodynamics effects~\cite{Przybytek2010,Hiyama2012b}, and many others~\cite{Hiyama2014}. These calculations consistently find that helium trimers and tetramers present universal aspects due to the underlying Efimov physics.

The $^4$He trimer is usually modeled by pairwise interactions with short-range repulsion and long-range attraction, while in this current work, we assume purely attractive two-body potentials and repulsive three-body forces. Although the interactions in helium systems differ from those employed in this study, some comparisons can be made concerning the universal window due to the scales involved. At the two-body level, the bond length of the $^4$He dimer, 52(4)~\AA~\cite{Grisenti2000}, is one order of magnitude larger than the effective range of 7.3~\AA~\cite{Janzen1995} and the van der Waals length of 5~\AA~\cite{Braaten2003}, thus universality is expected. Although the bond length of the $^4$He trimer, 11(5)~\AA~\cite{Bruhl2005}, exceeds the effective range and van der Waals length, it is of the same order as both. We observed similar behavior in our results for the universal window since the values of $\sqrt{\langle r^2\rangle}/r_0$ range from $\sim$ 6 to 20. Using the values of the scattering length and binding energies reported in Ref.~\cite{Kunitski2015}, we arrive at $x=0.11$ and $y=0.76$ for the ground state, and $x=0.75$ and $y=0.11$ for the first excited state. Although these values are not contemplated simultaneously in this study, this suggests that other potentials could have wider universality windows than those employed in this work.

\subsection{Universal trimers}

\begin{figure}[!htb]
\centering
\includegraphics[angle=-90,width=\linewidth]{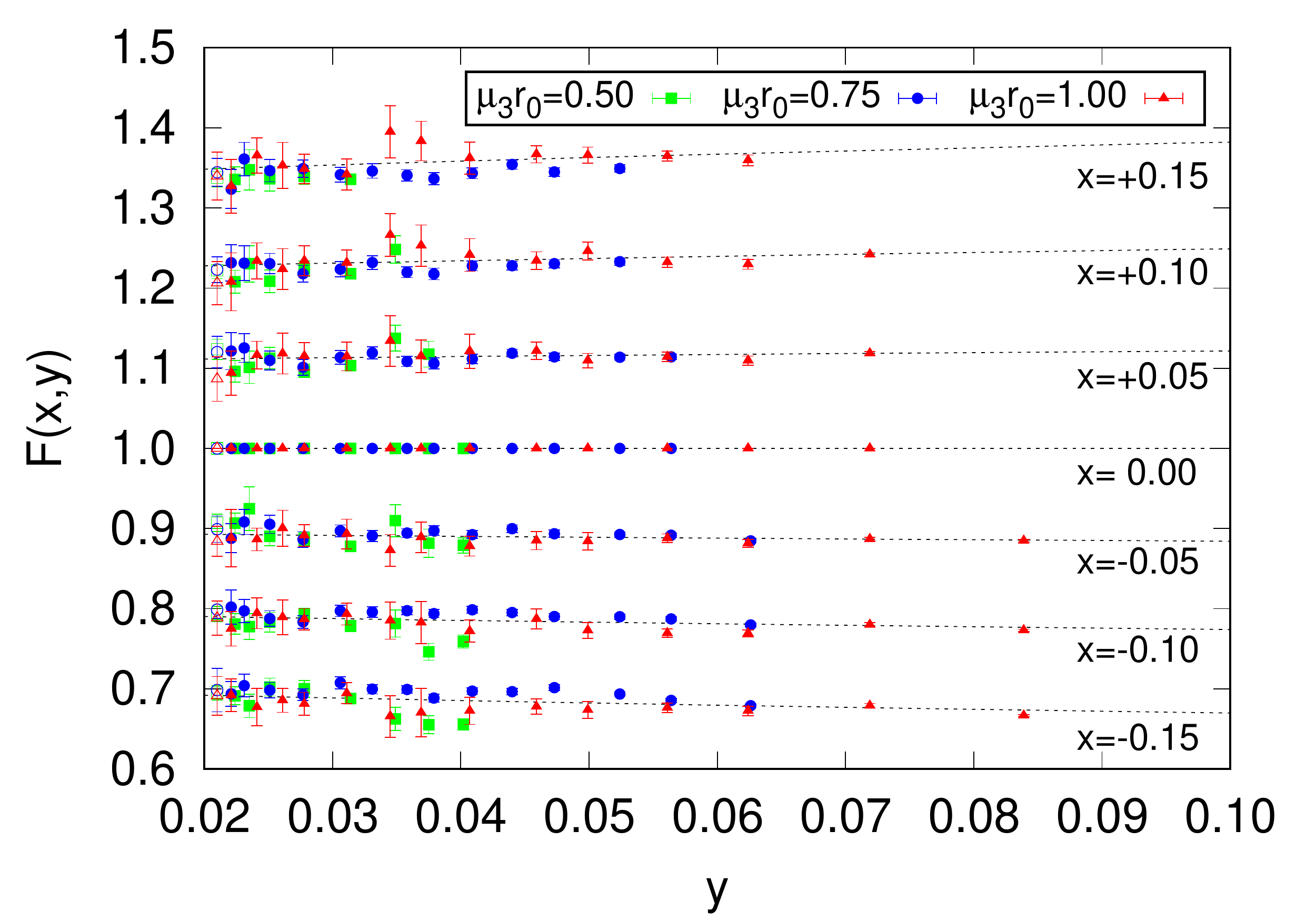}
\includegraphics[angle=-90,width=\linewidth]{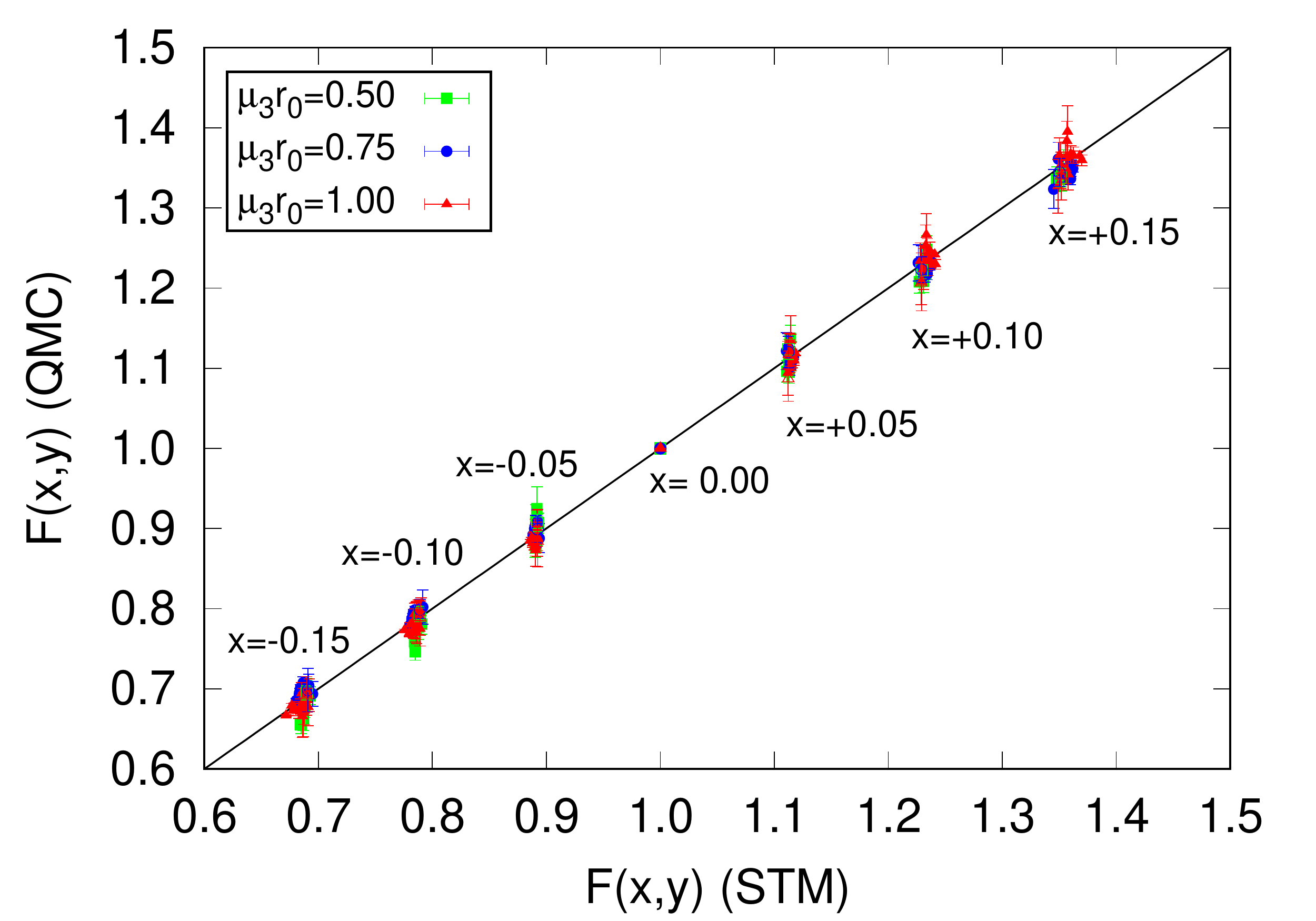}
\caption{(Color online) Upper panel: comparison between the scaling function obtained with the subtracted form of the STM equation (dashed curves) and the microscopic Hamiltonian with Gaussian two- and three-body potentials (full symbols) after the criterion $1/(\mu_3 R_3)\leqslant 0.08$ has been applied. The open symbols at the smallest value of $y$ considered correspond to calculations using the modified Poschl-Teller potential, Eq.~(\ref{eq:mPT}), for the two-body interactions.
Lower panel: quantum Monte Carlo calculations for $F(x,y)$ against the limit cycle results for the same values of $x$ and $y$. 
}
\label{fig:cut}
\end{figure}

Finally, we take all the results of our simulations for $-0.15\leqslant x\leqslant 0.15$, shown in Fig.~\ref{fig:all}, apply the criterion $1/(\mu_3 R_3)\leqslant 0.08$, depicted in Fig.~\ref{fig:mu3_r2}, to select the trimers that are in accordance with the scaling function. We display them in the upper plot of Fig.~\ref{fig:cut} where, besides the results for the Gaussian two-body potential, Eq.~(\ref{eq:gauss}), we present the ones obtained with the modified Poschl-Teller potential, Eq.~(\ref{eq:mPT}), for the smallest $y$ value.
Statistical fluctuations are present and, as seen in the plot of Fig.~\ref{fig:all}, the error tends to increase toward smaller values of $y$.
Still, it is possible to see the agreement between the results using microscopic Hamiltonians and the scaling function obtained with the subtracted form of the STM equation.

In particular, by increasing $y$, namely, by enhancing the trimer binding, and within the universal window, the dependence of $F(x,y)$ with $y$ is explored. Although showing small effects for $y\to 0.1$ as shown in Fig.~\ref{fig:all}, the smaller error in the trimer energies clearly corroborates the limit cycle found for this scaling function. That is illustrated by the Gaussian potential calculations with three-body forces with the smallest range $\mu_3 r_0=1$.

The physical information of the effective range for these universal trimers comes from a region in configuration space where the bosons are well separated, namely two of the bosons interact virtually and the spectator boson perceives only the asymptotic wave function of the pair interaction, which is contained in the on-shell scattering amplitude.
Furthermore,
the effective range enhances the attraction between the bosons for $a>0$ with respect to the unitary limit, while for $a<0$ by increasing $y$ the effect is repulsive, thus decreasing the trimer binding with respect to the unitary limit. Such physical effects are also supported by the Monte Carlo results. 

The plot shown in the lower panel of Fig.~\ref{fig:cut} gauges the quality of the comparison between the Monte Carlo results for $F(x,y)$ and the corresponding values obtained in the STM calculations. The results are clustered and associated with each $x$ value.
For smaller $y$ the statistical error systematically increases, as a consequence of the necessity of a stronger three-body repulsion to fit the trimer in the universal window.

\section{Conclusions and Outlook}
\label{sec:conclusions}

Our goal was to determine an energy scaling function for bosonic trimers close to the unitarity limit (infinite scattering length), which considers finite range effects. The scaling function was determined by considering the subtracted form of the STM equation with a leading order effective range correction. In addition, we validated the scaling function by comparing it to QMC calculations with microscopic two- and three-body potentials. We employed Gaussian and modified Poschl-Teller two-body potentials with a Gaussian three-body interaction. 

Concerning the calculations using the subtracted form of the STM equation, it is interesting to see that the limit cycles survive the addition of the finite range, see Fig.~\ref{fig:limitcycle}, which was not apparent beforehand. The physics behind the two-body contact parameter is also captured in the scaling function through the linear term in $x$. It would be very interesting to see if the other terms could be similarly related to observables. 

We should remark that all the results obtained with QMC methods concern the trimer ground state. This represents an alternative to the standard approach of probing Efimov physics through excited states.

For convenience, we chose the reference energy to be the energy of the trimer at unitary. Inspection of Eq.~(\ref{eq:universal}) reveals that we could choose the energy at some other scattering length, and the scaling would also be possible. This is relevant for experiments where different scattering lengths can be observed, but the unitary regime may be out of reach.

A few words to put our results in the context of previous related works are in order. The zero-range hyper-radial equation for the trimer, written down by Efimov, contains a term proportional to $1/R^2$. Efimov noted that range corrections would introduce a term proportional to $r_0/R^3$ \cite{Efimov1991}. This hyper-radial equation is discussed in detail in Ref.~\cite{Platter2009}, in the context of EFT. The authors derived a term proportional to $r_0/R^3$ as a perturbation and obtained corrections linear in the two-body effective range on the three-boson bound-state spectrum when $|a|\gg r_0$. The nonlinear behavior in the effective range that we found in this work, namely the $r_0^\sigma$ terms in Eq.~(\ref{eq:F}), cannot be obtained with linear corrections in $r_0$.
Its validity is confirmed by the agreement of the STM and QMC results for relatively large values of $y$; see the points with $y\gtrsim 0.07$ in Fig.~\ref{fig:cut}.

We should also mention the studies of Refs.~\cite{Kievsky2014,Gattobigio2014,Kievsky2015}, where two-body potentials have been used to derive range corrections to bosonic clusters. The main difference in our approach is that by including a repulsive three-body force, we can probe much smaller values of $y$, while these other works are in the region $y\sim 0.7$. In this way, both approaches are complementary. It would be interesting to derive a formal route from one limit to the other, but this is beyond the scope of the current work.

In this work, we considered particles with equal masses and the same interactions between them. It would be interesting to apply the same framework to mass-imbalanced systems~\cite{Bellotti2013,Yamashita2013,Sandoval2016,Rosa2019}.
Moreover, if we consider interactions with different scattering lengths between the pairs, we would have three instead of a single $a$ value and also different effective ranges.
Investigating if a universal scaling exists in this more complicated setting could be helpful to describe systems in atomic and nuclear physics. 

We centered our discussion around three particles, but the QMC methods employed in this work have been applied to bulk matter and clusters of up to 60 bosons~\cite{Carlson2017}. Even in the case of interactions chosen to reproduce relatively shallow trimers, finite-range effects appear for 15 particles or more~\cite{Stecher2010,Nicholson2012,Kievsky2014,Carlson2017}.
This happens because the interparticle spacing decreases, and the
range of the two- or three-body interactions become significant.

We intend to investigate if it is possible to construct analogous scaling functions to the one in this work for $N$-boson systems. We hope that the finite range effects of the interactions in these cases can restore what has been dismissed as nonuniversal behavior for large bosonic clusters.

\begin{acknowledgments}
We thank V.S. Bagnato for the useful discussions.
This work was partially supported by S\~ao Paulo Research Foundation (FAPESP)
grants 2018/09191-7 (L.M.), 2017/05660-0 (T.F. and L.T.), and 2019/00153-8 (M.T.Y.),
and Conselho Nacional de Desenvolvimento Cient\'{\i}fico e Tecnol\'{o}gico 
grants 308486/2015-3 (T.F.), 304469-2019-0(L.T.), and 303579/2019-6 (M.T.Y.). 
The work of S.G. was supported by U.S.~Department of Energy, Office of Science,  Office of Nuclear Physics, under Contract No.~DE-AC52-06NA25396, by the DOE NUCLEI SciDAC Program, by the LANL LDRD Program, and by the DOE Early Career Research Program.
The work also used the Extreme Science and Engineering Discovery Environment (XSEDE) Stampede2 
through the allocation TG-PHY160027, supported by National Science Foundation 
grant number ACI-1548562.
\end{acknowledgments}

\appendix

\section{ Ansatz for the scaling function}
\label{sec:appendix}

In our analysis, we are concerned with a very small region near unitarity, where we want to add range corrections. We found it convenient to express the three-body energy ratio in terms of the first terms of an expansion, instead of the exponential form presented in Ref.~\cite{Braaten2006}, $E_3(1/a,0,\nu)=E_2+E_{3,0}(0,0,\nu) e^{{\Delta(\eta)}/{s_0}}$, in which the corrections are within the factor $\Delta(\eta)$, where $\tan\eta=
\sqrt{E_2/E_{3}(0,0,\nu})$. For that, with $E_3\equiv E_3(1/a,0,\nu)$, $E_{3,0}\equiv E_3(0,0,\nu)$, we define $\chi$ and $\chi_0$ such that
\begin{equation}\label{chir0}
\chi\equiv\sqrt{\frac{E_2}{E_3}}\;\;{\rm and}\;\;
\chi_0\equiv\sqrt{\frac{E_2}{E_{3,0}}}. 
\end{equation}
The value of $\chi_0$, at the zero-range limit, can be identified with $x$ of Eq.~\eqref{eq:x}. Our corresponding fitted expression, obtained from systematic STM zero-range calculations, by keeping terms only up to the order of the two-body energy, is given by
\begin{equation}
F(x,0)=\frac{E_{3}}{E_{3,0}}\simeq
1+{2.107}\chi_0+0.804\chi_0^2.
\end{equation}
This ratio of trimer energies, out and at unitarity, is the convenient starting expression to introduce the effective range contribution. For that, by considering Eq.~\eqref{aB} we move to a range-dependent expression with $a\to a_B$ as suggested, such that 
$\chi_0\to\sqrt{{E_B}/{E_3(0,r_0,\nu)}}=x\left(1+{xy}/{2}\right)$ $=x\left(1+r_0/{2a}\right)$:
\begin{equation}\label{eq:E3ar2}
F_0(x,y)\simeq
1+x\left(1+\frac{xy}{2}\right)\left[{2.107}+0.804 x\left(1+\frac{xy}{2}\right)\right].\nonumber
\end{equation}
We can drop the last term inside the square brackets as it is of the order $1/a^3$. However, by using the procedure described above, we are still carrying the three-body energy at unitarity. As the effective range is also directly associated with the energy scale~\cite{Ji2015}, we can translate it to a shift in the energy scale, which can be introduced by an extra factor multiplying $\chi_0$, which we assume is given by 
$\left(1+ \frac{2s_0}{\pi}y^{\frac{2s_0}{\pi}}\right)=\left(1+ 
\frac{2s_0}{\pi}\exp({\frac{2s_0}{\pi}}\ln y)\right)$.
Note that the range correction is such that $\ln y < 0$, with this 
exponential factor varying between 0 and 1, such that the following scaling function is obtained: 
\begin{eqnarray}\label{fxy0}
    F(x,y)&=& \frac{E_3(1/a,r_0,\nu)}{E_3(0,r_0,\nu)}\\ 
&\simeq&1+2.107x+1.350xy^{\frac{2s_0}{\pi}}+0.804x^2\nonumber\\
&+&1.030x^2y^{\frac{2s_0}{\pi}}+1.053x^2y+ \cdots 
.
\nonumber\end{eqnarray}
If we consider large scattering lengths and set $y=0$, this expression is equivalent to Eq.~(\ref{eq:braaten0}).
We should observe that this parametrization is suitable for positive $y$ and shown to be adequate for analyzing the trimer energies obtained with the models used here.

Our motivation to prescribe the above ansatz is based on previous works 
(see Refs.~\cite{Kievsky2015,Gattobigio2019}), which have shown that the effective range correction can be translated to a shift in the energy scale with the effective range~\cite{Ji2015}. This was the starting point to build the approximate formula given by Eq.~\eqref{eq:F}. 
Besides that, a physical constraint to be added is that this shift must reflect the situation in which the three-body scale is no more relevant for the physics of the system, as in the case of a system formed by two light bosons and a very heavy one. This could occur for $s_0\ll 1$, i.e., when the separation between the energies of consecutive Efimov states 
tends to infinity, such that the three-body scale is no more relevant 
for the system. By following this reasoning, the sensitivity to the change in the short-range scale is reduced. At the same time, the $y$ dependence disappears with the exponent being proportional to $\sim s_0$. The natural coefficients are of $\mathcal{O}\sim 1$, both for the proportionality constant as well as for the exponent. Furthermore, for the Efimov discrete scaling, the relevant factor is $s_0/\pi$.

\bibliography{/home/madeira/backup/mendeley/bibtex/3bosons.bib}
\end{document}